\documentclass{sig-alternate}
\usepackage{graphicx}
\usepackage{acronym}
\usepackage{subfigure}
\usepackage{booktabs}
\usepackage{algorithm}
\usepackage{algcompatible}
\usepackage{url}
\usepackage{amstext}
\usepackage{amsmath}

\acrodef{P2P}{Peer-to-Peer}
\acrodef{TTL}{Time-To-Live}

\begin{document}
 
%
%
\title{Searching in Unstructured Overlays Using Local Knowledge and Gossip}
\numberofauthors{1}
\author{
\alignauthor
Stefano Ferretti\\
       \affaddr{Department of Computer Science, University of Bologna}\\
       \affaddr{Bologna, Italy}\\
       \email{sferrett@cs.unibo.it}}

\maketitle              

\abstract*{
This paper analyzes a class of dissemination algorithms for the discovery of distributed contents in Peer-to-Peer unstructured overlay networks. The algorithms are a mix of protocols employing local knowledge of peers' neighborhood and gossip. By tuning the gossip probability and the depth $k$ of the $k$-neighborhood of which nodes have information, we obtain different dissemination protocols employed in literature over unstructured P2P overlays. The provided analysis and simulation results confirm that, when properly configured, these schemes represent a viable approach to build effective P2P resource discovery in large-scale, dynamic distributed systems.  }

\abstract{
This paper analyzes a class of dissemination algorithms for the discovery of distributed contents in Peer-to-Peer unstructured overlay networks. The algorithms are a mix of protocols employing local knowledge of peers' neighborhood and gossip. By tuning the gossip probability and the depth $k$ of the $k$-neighborhood of which nodes have information, we obtain different dissemination protocols employed in literature over unstructured P2P overlays. The provided analysis and simulation results confirm that, when properly configured, these schemes represent a viable approach to build effective P2P resource discovery in large-scale, dynamic distributed systems.  }

\section{Introduction}

This paper deals with resource discovery in large-scale, dynamic \ac{P2P} distributed communication systems. In this context, it has been recognized that an interesting approach consists in exploiting unstructured overlay networks \cite{Cholvi:2004,simplex,Keidar:2006}, which are alternative to traditional structured solutions \cite{Hidalgo:2011}. 
Indeed, there are some clear drawbacks related to unstructured networks, that make structured ones more effective in some distributed systems. In particular, the main weakness of unstructured nets is that links among nodes do not depend on the distribution of the contents. 
This means that in general it is not possible to provide a bound on the number of nodes that might be involved during the lookup of a resource.
On the other hand, the advantages are the easier manageability and the possibility of implementing resource discovery systems based on partial-match and complex queries. Conversely, several structured P2P approaches (e.g.~those based on DHTs) strongly limit the expressiveness of the queries to retrieve contents. 
For these reasons, understanding if, how and when unstructured overlays can support resource and content lookup represents an interesting research topic.

A main aspect refers to the algorithm employed to distribute queries among nodes, that strongly influences the performance of the whole system.
In this paper, we study a simple class of dissemination algorithms, which are a mix of push-gossip based and informed propagation schemes \cite{disio11}. Each node has knowledge of its $k$-neighborhood, i.e.~those nodes that are distant at most $k$ hops from it. 
This information is exploited during the routing of messages in the overlay, i.e.~a node sends the message to those $1$-neighbors that can relay the message to the $k$-neighbours that hit the query. Moreover, the node gossips the message to its remaining $1$-neighbors.
The tuning of the parameters of the algorithm (i.e.~gossip probability threshold and depth $k$ of the $k$-neighborhood) allows to pass, for instance, from pure locally ``best neighbor selection'' dissemination protocols (gossip probability set equal to $0$), e.g.~\cite{PuttaswamySZ08}, to flooding schemes (gossip probability set equal to $1$). Similarly, if the depth $k$ of the $k$-neighborhood is set $k=0$, a pure gossip strategy is obtained; when $k$ is set equal to the network diameter, we have a scheme with full-knowledge of the net.

We present an analytical framework that models the described family of communication protocols. A numerical analysis over scale-free network topologies is performed, and it is compared with a simulation of the system. 
Results confirm that dissemination protocols exploiting the combination of gossip and local knowledge about nodes' neighborhood, are a useful tool to build lookup discovery services over large-scale unstructured P2P systems. 
Moreover, the framework can be practically exploited to tune the gossip probability at peers and build effective lookup discovery services over P2P unstructured overlays.
In many cases, it is sufficient to maintain information on the $2$-neighborhood (or even $1$-neighborhood, with a higher gossip probability) to have that queries percolate through the overlay, hence obtaining a number of query hits of the order of the number of resources (matching the query) present in the network.

The remainder of this paper is organized as follows. Section \ref{sec:model} presents the system model and the local protocol executed at each node. Section \ref{sec:cn} presents the mathematical model. Section \ref{sec:exp} outlines results coming from numerical analysis and simulation. Finally, Section \ref{sec:conc} provides some concluding remarks.

\section{System Model and Protocol}\label{sec:model}

Let consider unstructured overlay networks, with peers that connect each other through a pseudo-random attachment process which shapes the overlay based on a specific network topology, defined through a degree probability distribution. The link creation process does not depend on the placement of contents in the P2P system \cite{ferretti_trans.cs.2012.10-12.e2}.
We denote with $\Pi^1$ the $1$-neighbor\-hood of a node $n$ ($n$'s friends); in general $\Pi^k$ is the $k$-neighborhood of a node, i.e.~nodes at most $k$ hops away from $n$. 
Nodes know how to reach all its $k$-neighbors. We assume the existence of a \textsc{relay}($m$) procedure that returns the node that $n$ has to contact to reach $m$. Of course, if $m$ is a $1$-neighbor of $n$, \textsc{relay}($m$) returns $m$.

When a peer $n$ holds (removes) from its cache a novel resource item, it informs its $k$-neighborhood, through some multicast message sent through the overlay.
Hence, upon reception at $m$ of a message stating that $n$ holds (deletes) a novel resource item, $m$ adds (removes) a related entry in its neighbor table. 
This way, each time $m$ receives a query that hits that resource item, $m$ can forward the query towards $n$.
It is clear that the higher the depth $k$ of the neighborhood, the higher the amount of control messages to be transmitted to maintain correct information.

The distribution of a query is based on pure local decisions \cite{disio11}. 
We assume that each query contains all the information needed to perform a matching among the requested (type of) item and resources available in the system; in other words, resources are described through a profile (or some metadata).
Algorithm \ref{alg:protocol} shows the pseudo-code of the peer ($n$) behavior executed to disseminate a query. When $n$ creates or receives a novel query from a neighbor $m$ (which has not be handled already, lines \ref{alg:b_handled}--\ref{alg:e_handled}), first, it checks if there is a query hit locally; in this case, the query originator is contacted directly (lines \ref{alg:b_queryhit}--\ref{alg:e_queryhit}). 

Then, $n$ multicasts the query to those $k$-neighbors that own an item that hits the query (lines \ref{alg:b_f}--\ref{alg:e_f}). This is accomplished by sending the message to its $1$-neighbors that will relay it to the target nodes. However, this is done only if the message has a positive \ac{TTL} (lines \ref{alg:b_ttl}--\ref{alg:b_ttl_check}). (We are assuming that the TTL value allows to cover the whole network; typically, this can be obtained using low values of the order of the logarithm of the network size.)
Finally, $n$ gossips the message with a probability $\gamma \leq 1$  to the remaining set of $1$-neighbors (lines \ref{alg:b_gossip}--\ref{alg:e_gossip}) \cite{disio11}.

\begin{algorithm}[t]
\caption{Query distribution protocol executed at node $n$}
\label{alg:protocol}
\begin{small}
\begin{algorithmic}[1]
\REQUIRE{Query $Q$ generated at $n$ $\vee$ $Q$ received in a message relayed by a neighbor peer $m$}
\IF{$Q$ already handled} \label{alg:b_handled} 
  \STATE Return
\ENDIF \label{alg:e_handled}
\IF{\textsc{QueryHit}($Q$)} \label{alg:b_queryhit}\hfill\COMMENT {local query hit} 
  \STATE $s$ = \textsc{Originator($Q$)}
  \STATE $rp$ = \textsc{ProfileMatchingResource($Q$)}
  \STATE $msg = \langle \text{``available''}, rp \rangle$
  \STATE \textsc{send}($msg, s$)%
\ENDIF \label{alg:e_queryhit}
\STATE \textsc{decreaseTTL}($Q$) \label{alg:b_ttl} 
\IF{\textsc{TTL}($Q$) $> 0$} \label{alg:b_ttl_check}\hfill\COMMENT {relay to hitting nodes}
 \STATE $R \leftarrow \{ \text{\textsc{relay}}(i) | i \in \Pi^k \wedge i$  has an item matching $Q\} \setminus m$ \label{alg:b_f}
 \FORALL{$r \in R$}
   \STATE \textsc{send}($Q$, $r$)%
 \ENDFOR \label{alg:e_f}
 \FORALL{$i \in \Pi^1 \setminus \{ R \cup m \}$ } \hfill\COMMENT {gossip}\label{alg:b_gossip}
     \IF{\textsc{random()} $< \gamma$}  %
         \STATE \textsc{send}($Q,i$)%
     \ENDIF%
 \ENDFOR \label{alg:e_gossip}
\ENDIF \label{alg:e_ttl}
\end{algorithmic}
\end{small}
\end{algorithm}

\begin{figure}[thbp]
   \centering
   \includegraphics[width=.6\linewidth]{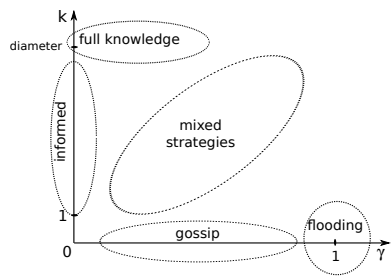}
  \vspace{-0.2cm}
\caption{Discovery protocols obtained through the setting of the depth of the $k$-neighborhood and the gossip probability $\gamma$.}
   \label{fig:prots}
\end{figure}

The considered family of protocols groups together different typical schemes employed over unstructured P2P overlays.
Figure \ref{fig:prots} shows the protocols we obtain depending on the gossip threshold $\gamma$ and depth of the $k$-neighborhood. In fact, when $k=0$ and $\gamma>0$, we have a gossip protocol, i.e.~queries are randomly disseminated. When $\gamma=1$ we have a flooding protocol, i.e.~messages are relayed through all nodes' links.
Informed protocols are those where peers have knowledge of their $k$-neighborhood (without using gossip) \cite{PuttaswamySZ08}; they are thus placed on the $k$-axis, with $\gamma=0$.
Finally, if we ideally set the $k$ value equal to the network diameter, then we obtain full-knowledge schemes, where the overlay is exploited to route messages.

\section{System Analysis}\label{sec:cn}

The goal of this analysis is to estimate the average amount of query hits $\langle h \rangle$ that would occur, given an estimate of the resource popularity (i.e.~how much resources, that would hit the query, are distributed in the net) and a given degree distribution probability characterizing the unstructured overlay topology. 

Each query dissemination process is considered as a standalone, independent task. This is a correct assumption if peers have a buffer cache whose size is sufficiently large to handle simultaneous queries. Otherwise, the model should be extended to consider possible buffer overflows. 

We assume to work with very large and dynamical P2P systems. We already mentioned that, for small-sized and stable nets, the use of unstructured overlays can be avoided, since other approaches can be proficiently employed, such as centralized solutions or structured distributed systems (e.g.~DHTs).
The high number of nodes, together with the random nature of contacts among peers in the overlay, augments the probability of having a low clustering in the network \cite{simplex,newmanHandbook}. 
A consequence of the random nature of the attachment process is that, regardless of the node degree distribution, the probability that a $2$-neighbor is also a $1$-neighbor of a node, goes as $N^{-1}$, being $N$ the number of nodes in the overlay. Hence, this situation can be ignored for high $N$ values. 
This assumption is supported by previous works, asserting that it is undesirable for an unstructured P2P overlay to have high clustering \cite{voulgaris.jnsm.2005}. In fact, clustering reduces the connectivity of a cluster to the rest of the net, increases the probability of partitioning, and it  may cause redundant message delivery. 


We denote with $p_i$ the probability that a peer has $i$ $1$-neighbors (its degree).
Let $q_i$ be the excess degree distribution \cite{newmanHandbook}, i.e.~the probability that, following a link in the overlay, we arrive to a peer $m$ that has other $i$ links (hence the degree of $m$ is $i+1$). Given $p_i$, we have that
$q_i = \frac{(i+1)p_{i+1}}{\sum_j j p_j}.$

Probabilities $p_i$ and $q_i$ represent two similar concepts i.e.~the number of contacts of a considered peer (its degree), and the number of contacts obtained following a link of a peer (its excess degree), respectively. In the following, we introduce measures obtained by considering the degree $p_i$ of a node, as well as the excess degree $q_i$ of a link. Hence, with a slight abuse of notation we denote all the probabilities/functions related to the excess degree with the same letter used for the degree, with an arrow on top of it, just to recall that the quantity refers to a link.
Thus, for instance, the generating functions for $p_i$ and $q_i$ are denoted as $ G(x) = \sum_i p_i x^i,  \overrightarrow{G}(x) = \sum_i q_i x^i$.

We denote with $\rho$ the probability that a node has a resource item matching the considered query, and with $\gamma$ the gossip probability.
If the considered protocol employs the $1$-neighborhood $\Pi^1$ only, then the probability that a node $n$ does not transmit a query to a neighbor $m$ is $(1-\rho)(1-\gamma)$, i.e.~the probability that $m$ does not hit the query, and $n$ decides not to gossip to $m$. Hence, the probability $\tau_1$ that $n$ transmits the query to a neighbour $m$, having only knowledge of its $1$-neighborhood $\Pi^1$ is $\tau_1 = 1 - (1-\rho)(1-\gamma)$.

With this in view, the probability that none of the $n$'s $1$-neighbours hit the query is $\sum_i p_i (1-\rho)^i = G(1-\rho)$. This result is obtained by considering all the possible cases of $n$ having degree $i$ and its $i$ neighbours do not hit the query. Similarly, the probability that, given a randomly chosen edge of $n$, we arrive to a node $m$ that does not have any neighbour (apart from the link we considered to arrive to $m$ from $n$) that hit the query is $\sum_i q_i (1-\rho)^i = \overrightarrow{G}(1-\rho)$. 

Following this reasoning, it is possible to determine the probability $\tau_2$ of relaying a query to a node $m$ when $n$ has knowledge of its $2$-neighborhood $\Pi^2$. In fact, such probability is 
$\tau_2 = 1 - (1-\rho)(1-\gamma)\overrightarrow{G}(1-\rho)$, i.e.~$n$ does not transmit to $m$ if: $m$ does not hit the query (probability $(1-\rho)$); $n$ decides not to gossip $m$ (probability $(1-\gamma)$); and $n$ knows that its $2$-neighbours connected through $m$ do not hit the query (probability $\overrightarrow{G}(1-\rho)$ measured above).

The approach can be exploited to measure $\tau_k$, with any given value of $k$. For instance, the probability that following a link we arrive to a node which has no neighbors in its $\Pi^2$ that hit the query is 
$\sum_i q_i (1-\rho)^i [\overrightarrow{G}(1-\rho)]^i = \overrightarrow{G}\big( (1-\rho){\overrightarrow{G}(1-\rho)} \big).$ 
Through this result we might obtain $\tau_3$, and so on.

Now, the probability that $n$ forwards a message to $i$ of its neighbors is
\begin{equation}\label{eq:f_k}
  f_i = \tau_k ^i \sum_{j \geq i} p_j \binom{j}{i} (1-\tau_k)^{j-i}.
\end{equation}
$f_i$ considers all the possible cases of $n$ having a degree $j$, which forwards the query to $i (<j)$ neighbors, while not forwarding the query to its remaining $j-i$ neighbors.
Similarly, the probability that following a link we arrive to a node that forwards the query to $i$ other nodes is readily obtained by substituting, in (\ref{eq:f_k}) above, $p_j$ with $q_j$, i.e.
$  \overrightarrow{f}_i = \tau_k^i \sum_{j \geq i} q_j \binom{j}{i} (1 - \tau_k)^{j-i}.$

If we consider the generating function $F$ of the $f_i$ coefficients, we have
\begin{eqnarray*}\label{eq:F_calcolo} 
  F(x) & = & \sum_i f_i x^i 
     =  \sum_i \tau_k^i x^i 
              \sum_{j \geq i} p_j \binom{j}{i} (1 - \tau_k)^{j-i}\nonumber \\
  & = & \sum_j p_j \sum_{i =0}^j \binom{j}{i}  \tau_k^i x^i (1 - \tau_k)^{j-i}\nonumber \\
   & = & \sum_j p_j (\tau_k x + 1 - \tau_k)^j \ 
   =\  G\big(\tau_k x + 1 - \tau_k\big).
\end{eqnarray*}
The average value of coefficients $f_i$ is given by the derivative of $F$ measured at $x=1$, i.e.~$F'(1) = \sum_i i f_i$, 
\begin{eqnarray*}\label{eq:F'(1)} 
  F'(x)\Bigl\lvert_{x=1} & = & \frac{dG}{dx}\big(\tau_k x + 1 - \tau_k\big)\Bigl\lvert_{x=1} 
 \ = \ \tau_k G'(1)\  
                         =\  \tau_k \langle p \rangle,
\end{eqnarray*}
where $\langle p \rangle$ is the mean node degree.

Similarly,,
$  \overrightarrow{F}'(x)\Bigl\lvert_{x=1}
  = \tau_k \overrightarrow{G}'(1) = \tau_k \langle q \rangle,$
where $\langle q \rangle$ is the mean value of the excess degree,
$\langle q \rangle =  \sum_i i q_i = \frac{\sum_i i (i+1) p_{i+1}}{\sum_j j p_j}
   =  \frac{\langle p^2 \rangle - \langle p \rangle}{\langle p \rangle}.$


With these measures, it is possible to obtain the whole number of nodes reached by a message starting from a given node, regardless of the number of hops \cite{newmanHandbook}. Let consider the probability $r_i$ that $i$ peers receive a query, starting from a given node and $\overrightarrow{r}_i$ is the probability that $i$ peers are reached starting from a link. $\overrightarrow{r}_i$ can be defined using the following recurrence,
\begin{align}\label{eq:r_k_l}
  \overrightarrow{r}_0 & =  0,\nonumber \\
  \overrightarrow{r}_{i+1} & =  \sum_{j \geq 0} \overrightarrow{f}_j \sum_{a_1 + a_2 + \ldots + a_j = i} \overrightarrow{r}_{a_1} \overrightarrow{r}_{a_2} \ldots \overrightarrow{r}_{a_j}.
\end{align}
Equation (\ref{eq:r_k_l}) can be explained as follows. It measures the probability that following a link we disseminate the query to $i+1$ peers. (The case $\overrightarrow{r}_0$ is impossible, since at the end of a link there must be a node.) One peer is that reached at the end of the link itself. Then, we consider the probability that the peer forwards to other $j$ links (varying the value of $j$). Each link $k$ allows to disseminate the query to $a_k$ peers, and the sum of all these reached peers equals to $i$.

Similarly, we can calculate $r_i$ as follows
\begin{align}\label{eq:r_k}
  r_0  &= 0,\nonumber \\
  r_{i+1}  &= \sum_{j \geq 0} f_j \sum_{a_1 + a_2 + \ldots + a_j = i} \overrightarrow{r}_{a_1} \overrightarrow{r}_{a_2} \ldots \overrightarrow{r}_{a_j}.
\end{align}
In this case, we start from the peer itself, considering it forwards to $j$ nodes; and as before, from these $j$ links we can reach $i$ other peers, in total.

The use of generating functions, $R(x) = \sum_i r_i x^i$, $\overrightarrow{R}(x) = \sum_i \overrightarrow{r}_i x^i$, allow to handle equations (\ref{eq:r_k_l}--\ref{eq:r_k}). In fact, after some algebraic manipulation we have
\begin{eqnarray}\label{eq:R_q_rec} 
  \overrightarrow{R}(x)  =  x \sum_{j \geq 0} \overrightarrow{f}_j [\overrightarrow{R}(x)]^j
     =  x \overrightarrow{F}(\overrightarrow{R}(x))
\end{eqnarray}
and, similarly,
\begin{eqnarray}\label{eq:R_rec} 
  R(x)  =  x \sum_{j \geq 0} f_j [\overrightarrow{R}(x)]^j
     =  x F(\overrightarrow{R}(x)).
\end{eqnarray}
From these generating functions, 
it is possible
to measure the average number $\langle r \rangle$ of peers that receive a query through the dissemination protocol, i.e.
$\langle r \rangle = \sum_i i r_i = R'(1)$. 
On the other hand, taking (\ref{eq:R_rec}) and differentiating 
\begin{eqnarray*}\label{eq:r} 
R'(1) = \big[F(\overrightarrow{R}(x)) + x F'(\overrightarrow{R}(x))\overrightarrow{R}'(x)\big]_{x=1}
 \  = \ 1 + F'(1)\overrightarrow{R}'(1),
\end{eqnarray*}
Similarly, from (\ref{eq:R_q_rec}),
$\overrightarrow{R}'(1) = \big[\overrightarrow{F}(\overrightarrow{R}(x)) + x \overrightarrow{F}'(\overrightarrow{R}(x))\overrightarrow{R}'(x)\big]_{x=1}
 \ = \ 1 + \overrightarrow{F}'(1)\overrightarrow{R}'(1).$
Thus,
$\overrightarrow{R}'(1) = \frac{1}{1 - \overrightarrow{F}'(1)},$
and final formula for $\langle r \rangle$ is
\begin{equation}\label{eq:r_final}
\langle r \rangle  =  1 + \frac{F'(1)}{1 - \overrightarrow{F}'(1)}
 =  1 + \frac{\tau_k\langle p \rangle^2}{(1+\tau_k)\langle p \rangle - \tau_k \langle p^2 \rangle}. 
\end{equation}

Now, $\langle r \rangle$ is the number of peers that receive the query, regardless if these nodes have a resource item matching it. To obtain the average number of query hits $\langle h \rangle$, it suffices to multiply $\langle r \rangle$ by the probability $\rho$ that a peer has a resource item matching that query, i.e.~$\langle h \rangle = \rho \langle r \rangle.$

Equation (\ref{eq:r_final}) has a divergence when $(1+\tau_k)\langle p \rangle = \tau_k \langle p^2 \rangle$, meaning that, under the assumption that the network has an infinite size, the query reaches an infinite number of nodes, i.e.~the query percolates through the network.
In other words, an amount of nodes of the order of the network size receives the query.

\section{Evaluation}\label{sec:exp}

This section presents an assessment performed by considering the analytical model and simulation. While during the assessment we tested different network topologies, we will focus here on results concerned with scale-free networks only.
These networks are characterized by nodes having a degree following a power law distribution $\sim p^\alpha$. 
They are characterized by the presence of hubs, i.e.~nodes with degrees significantly higher than the average, that have an important impact on the net connectivity.
The interest on scale-free networks in this work relates to the fact that several real \ac{P2P} systems are indeed scale-free networks \cite{simutools,newmanHandbook}.

In this study, we considered not only traditional scale free networks, but also those with an abrupt cutoff $c$ that limits the maximum degree that peers can maintain, so as to bound the workload that hubs in the P2P system must sustain. 

\subsection{Simulation}

We have built a discrete-event simulator mimicking the presented protocol. The simulator was written in C code and it allows testing the behavior of a set of nodes executing the presented dissemination protocol.
It is able to generate a random network based on a chosen degree distribution. In particular, once having (randomly) assigned a specific target degree to each node, using the selected degree distribution, a random mapping is made so that links are created until each node has reached its own target degree.
During the initialization phase, for each node a random choice was made to place resources; the resource availability was set based on a probability $\rho$, i.e.~for each network node, an item was present with probability $\rho$.

To build scale-free networks, the construction method was the one proposed in \cite{Aiello00arandom}.
This algorithm differs from other well known proposals, which build networks with a power law distribution by continuously adding novel nodes, hence having networks that grow in time. Conversely, we build a network of fixed size, characterized by two parameters $a, b$. 
More specifically, 
the number $y$ of nodes which have a degree $x$ satisfies $\log{y} = a - b \log{x}$, i.e.~$y = \lfloor\frac{e^a}{x^b}\rfloor$. 
Thus, the total number of nodes 
$N = \sum_{x=1}^{\lfloor e^{\frac{a}{b}}\rfloor} \frac{e^a}{x^b},$
being $\lfloor e^{\frac{a}{b}}\rfloor$ the maximum possible degree of the network, since it must be that $0 \leq \log{y} = a - b \log{x}$.
Once the number of nodes and their degrees have been determined, edges are randomly created among nodes until nodes reach their desired degrees.
In the reported results, the parameters were set to $a=6$, $b=1$, resulting in networks composed of $2482$ nodes. 

For each overlay, we varied the values of $\sigma, \rho$ in a range going from $0.01$ up to $0.5$, using a step of $0.01$. Thus, $2500$ simulation scenarios were considered. 
For each of these settings, we repeated the simulation using a corpus of $20$ different randomly generated networks (characterized by the mentioned statistical properties of the target topology). During each simulation execution, we analyzed the dissemination of $400$ queries sent by random nodes. 

\subsection{Results}

In a scale free network (without cutoffs) it is known that when $\alpha > -2$ the mean diverges; when $-3 < \alpha < -2$, the mean is finite but the variance and higher moments diverge \cite{newmanHandbook}. Hence, in these cases a query easily percolates through the network and resources are found with high probability. 
Indeed, results from our assessment confirm this. (We do not show them in charts.)

\begin{figure*}[thbp]
   \vspace{-0.5cm}
   \centering
   \subfigure[Model, $\Pi^1$]{\includegraphics[angle=-90,width=0.46\linewidth]{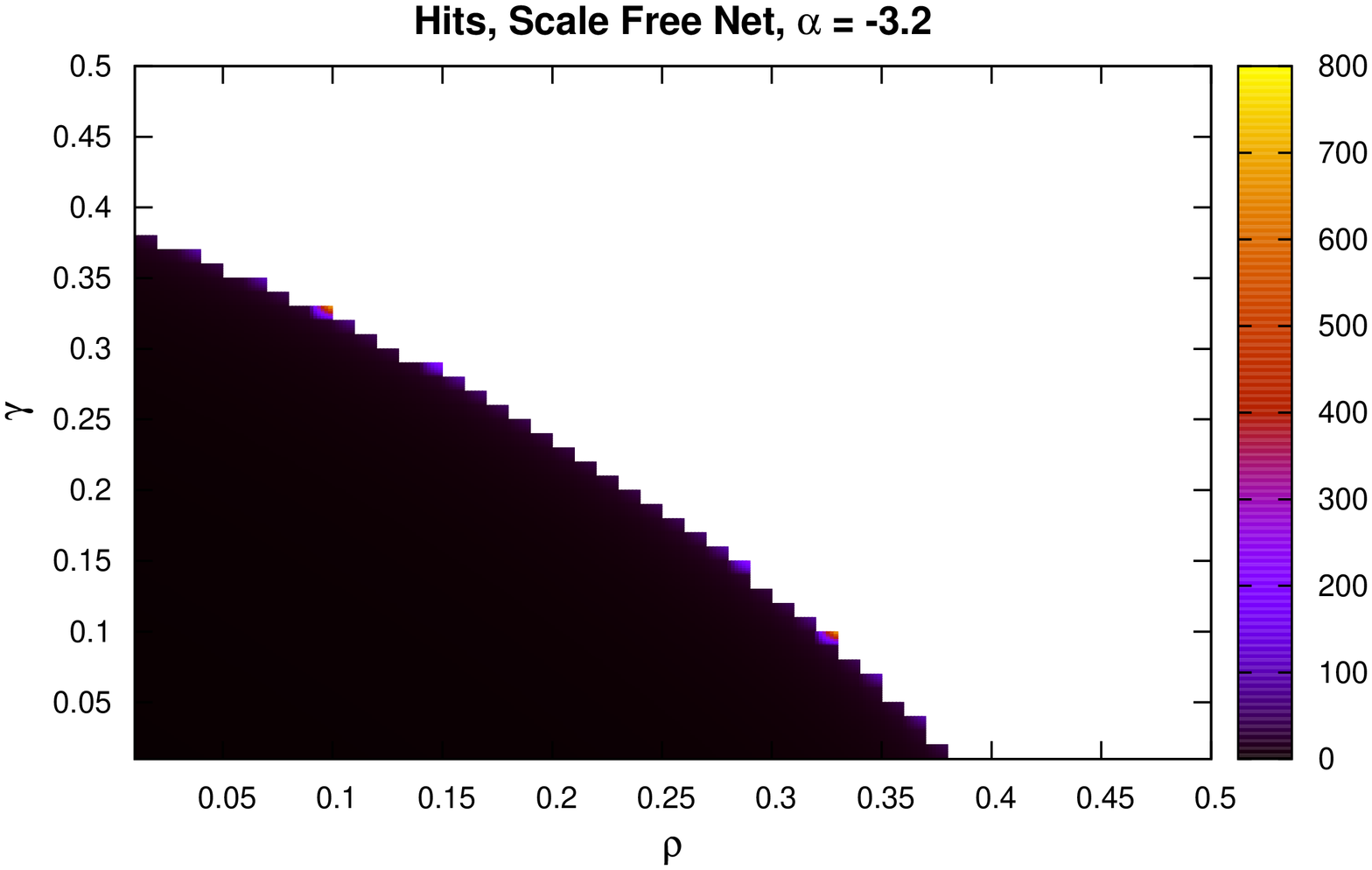}}
 \hspace{5mm}
   \subfigure[Simulation, $\Pi^1$]{\includegraphics[angle=-90,width=0.46\linewidth]{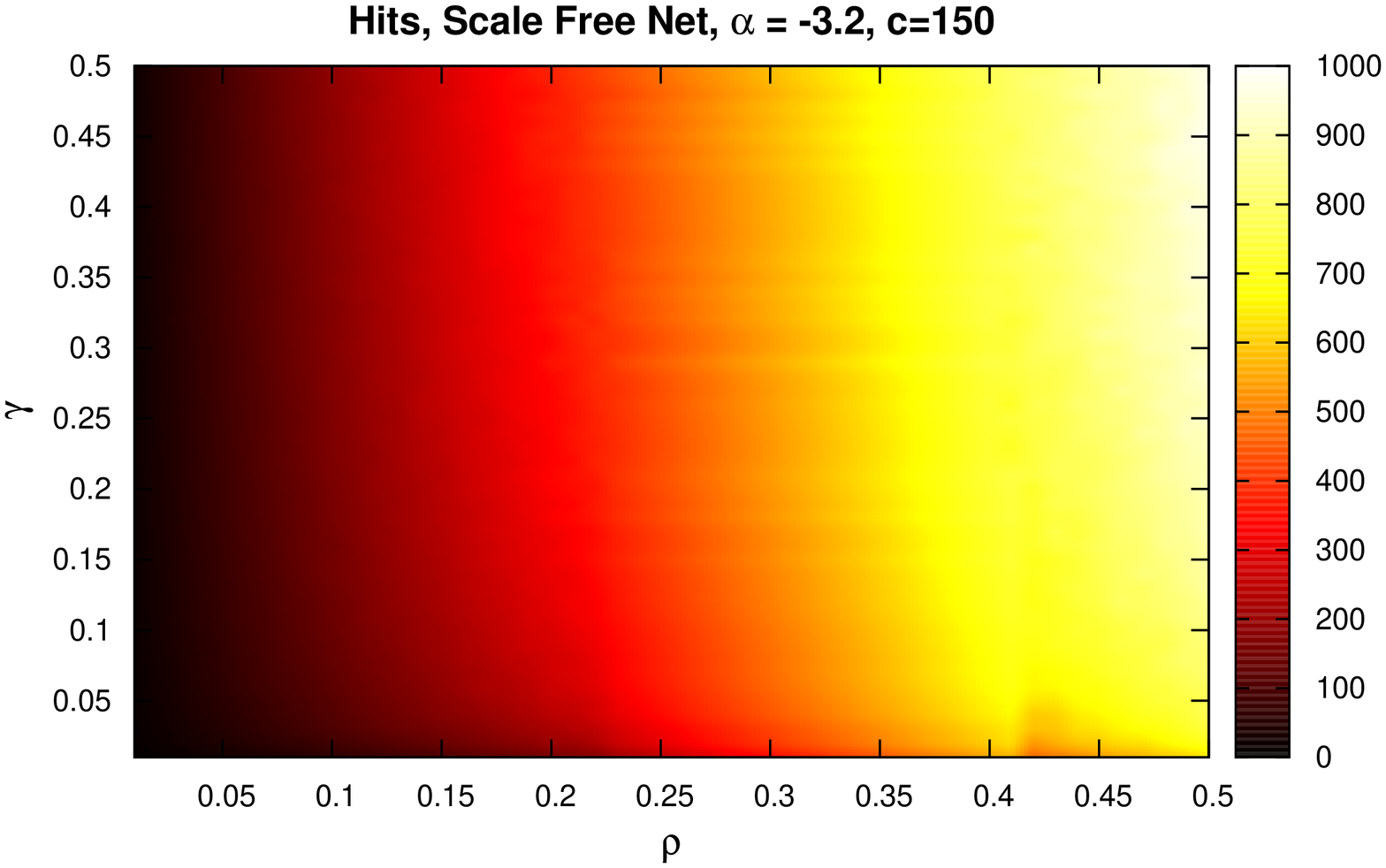}}
\caption{Average amount of query hits; power law degree distribution with exponent $\alpha=-3.2$. Results are shown for $\Pi^1$. When $\Pi^2$ is considered, the model returns an $\infty$ amount of query hits regardless of $\rho, \sigma$ values (hence not shown in the figure); simulation results confirmed that a high majority of nodes is reached and that queries percolate through the net.}
   \label{fig:sf_res}
     \vspace{-0.5cm}
\end{figure*}

For this reason we focus, for now, on overlays with a lower value for such exponent, i.e. $\alpha=-3.2$.
Figure \ref{fig:sf_res} shows the average amount of query hits in this specific scenario, obtained via the analytical model and simulation, when peers know their $1$-neighborhood $\Pi^1$. (In fact, when peers have knowledge of $\Pi^2$, the number of receivers diverges, and thus each query percolates through the network.)
It is possible to observe that with lower values of $\gamma, \rho$ a limited amount of network nodes receive the disseminated queries. Then, by increasing these two values, we reach a transition phase; and after that, the query percolates.
One might notice some differences between the two charts referring to the analysis and simulation. Actually, these are perfectly reasonable since the analysis assumes an infinite network size; hence, once a message percolates an infinite amount of nodes is reached. Conversely, simulations employed finite networks; hence, we obtain smoother transitions where a finite (nevertheless significant, when percolation occurs) amount of nodes is reached.
With this in view, we can conclude that the two approaches provide similar results.

\begin{figure}[thbp]
   \centering
   \includegraphics[width=.6\linewidth]{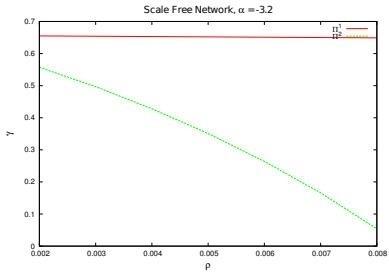}
\caption{Minimum $\gamma$ to find at least one resource; power law degree distribution with exponent $\alpha = -3.2$.}
   \label{fig:one_res_sf}
   \vspace{-0.2cm}
\end{figure}

Figure \ref{fig:one_res_sf} shows the minimum value of the gossip probability $\gamma$, to have that at least one resource is found through a query in a scale free network with $\alpha=-3.2$. The outcome has been obtained through a numerical analysis exploiting the mathematical model. When peers have knowledge of $\Pi^2$, with a resource presence probability $\rho > 0.008$ the gossip probability can be set $\gamma = 0$; hence, a non-negligible threshold for the gossip probability is needed only for rare items. This result is due by the presence of hubs that manage information of a high number of nodes.

\begin{figure*}[thbp]
   \centering
   \subfigure[$\alpha = -2.8$, neighborhood $\Pi^1$]{\includegraphics[width=0.47\linewidth]{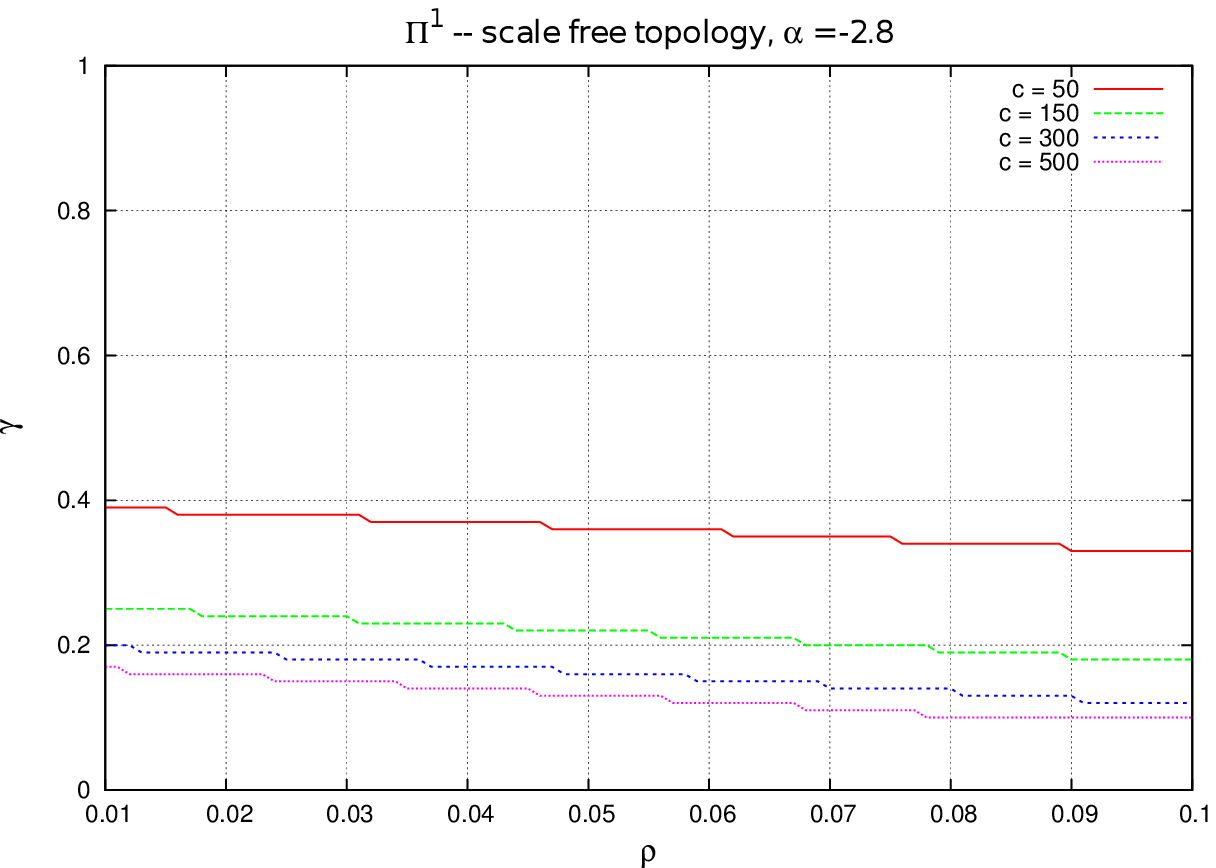}}
 \hspace{5mm}
   \subfigure[$\alpha = -2.8$, neighborhood $\Pi^2$]{\includegraphics[width=0.47\linewidth]{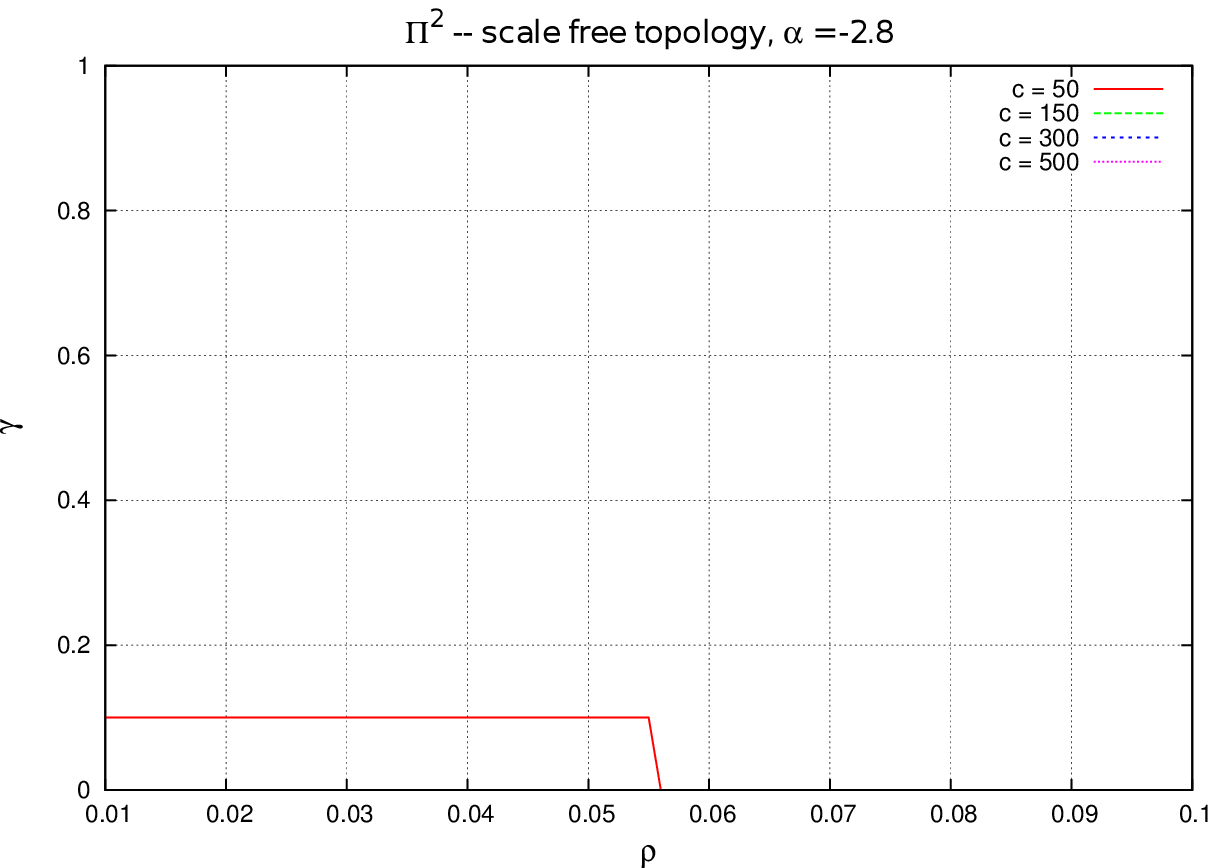}}
   \subfigure[$\alpha = -3$, neighborhood $\Pi^1$]{\includegraphics[width=0.47\linewidth]{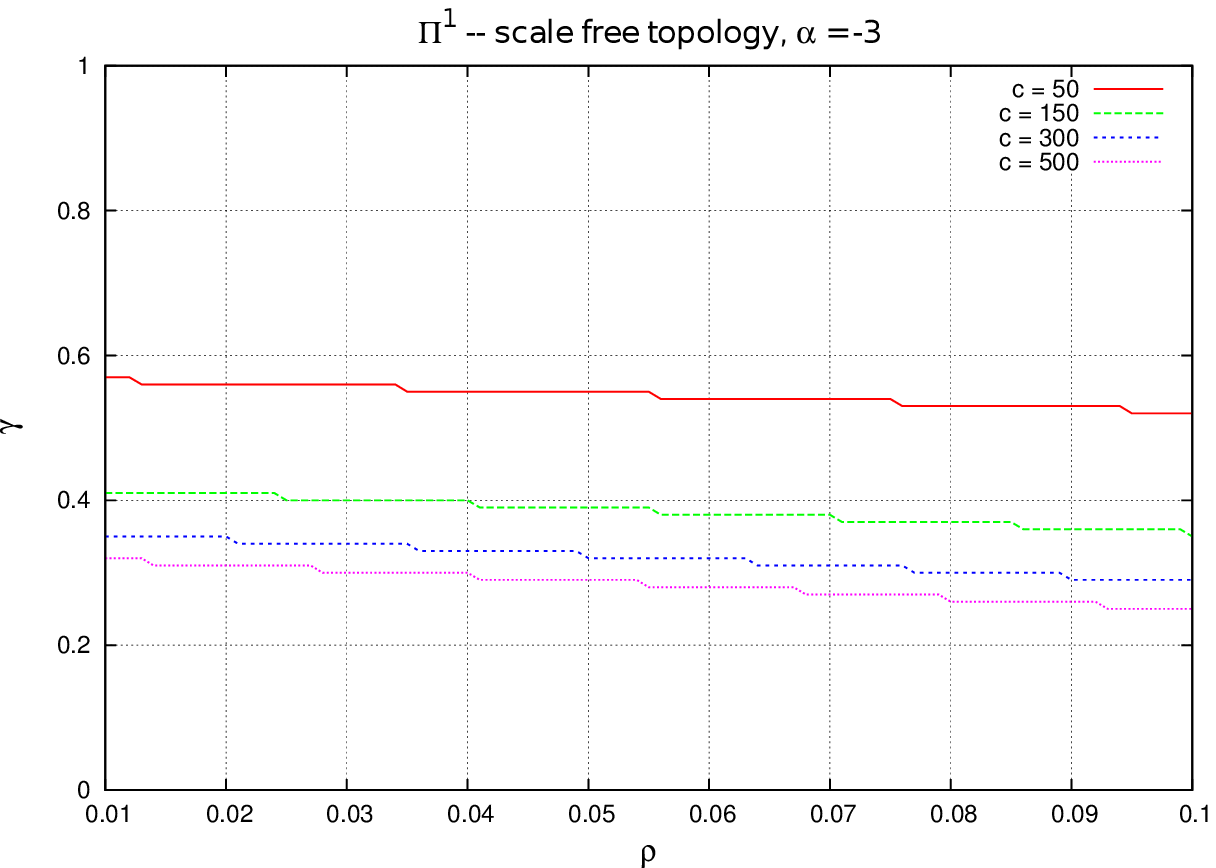}}
 \hspace{5mm}
   \subfigure[$\alpha = -3$, neighborhood $\Pi^2$]{\includegraphics[width=0.47\linewidth]{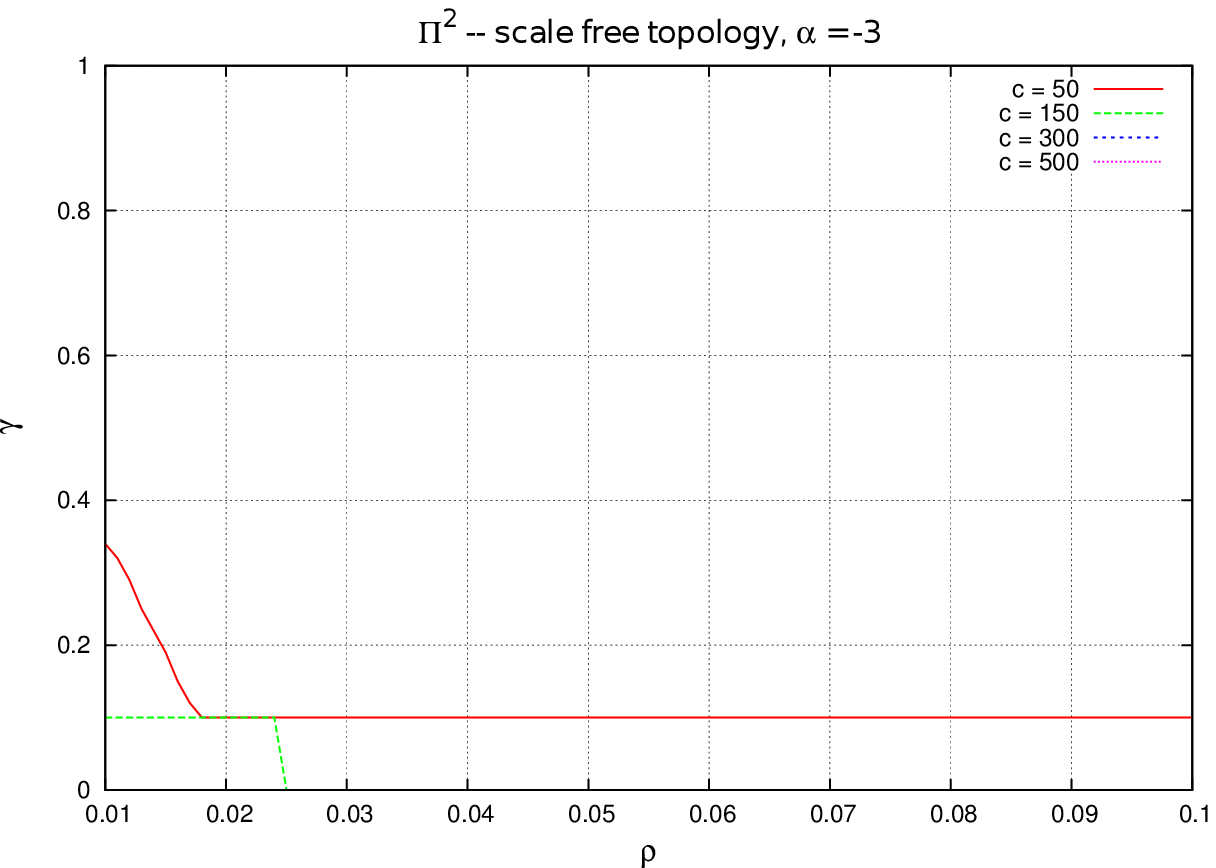}}
   \subfigure[$\alpha = -3.2$, neighborhood $\Pi^1$]{\includegraphics[width=0.47\linewidth]{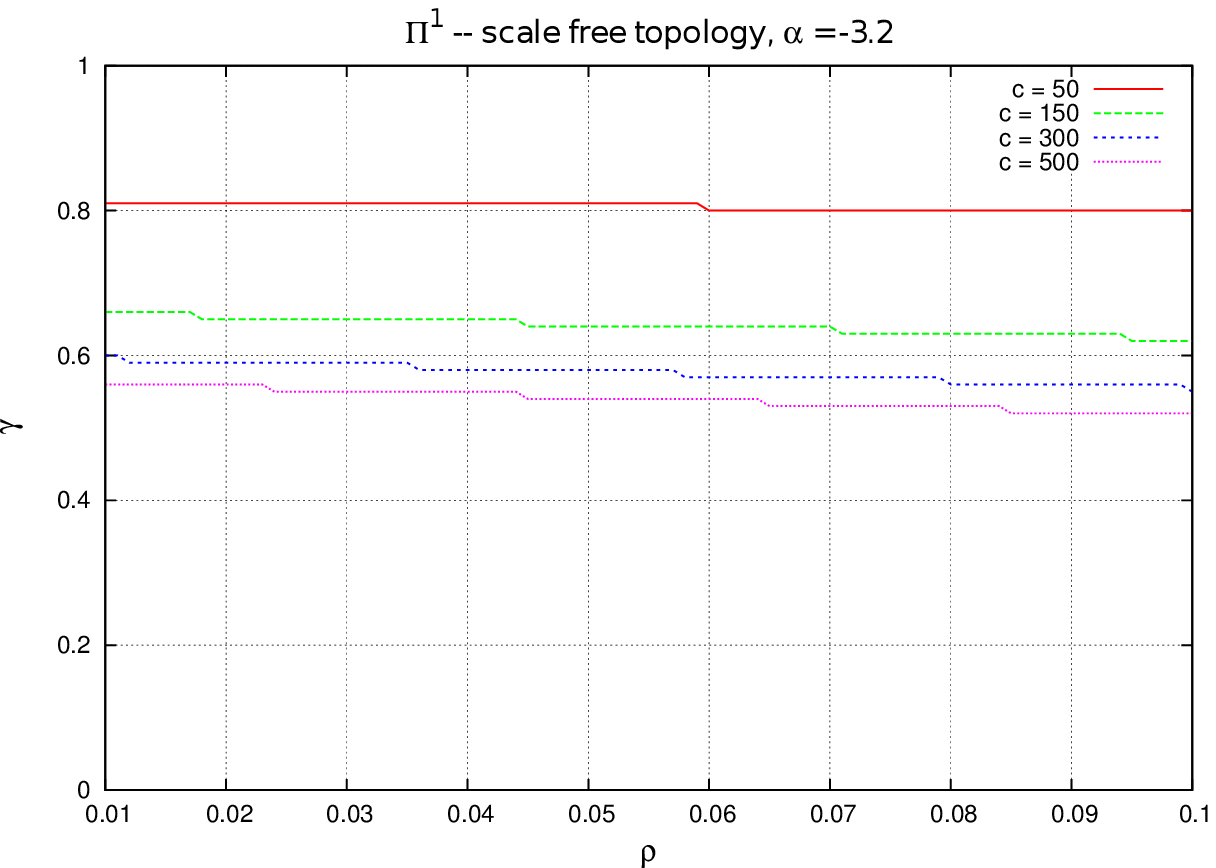}}
 \hspace{5mm}
   \subfigure[$\alpha = -3.2$, neighborhood $\Pi^2$]{\includegraphics[width=0.47\linewidth]{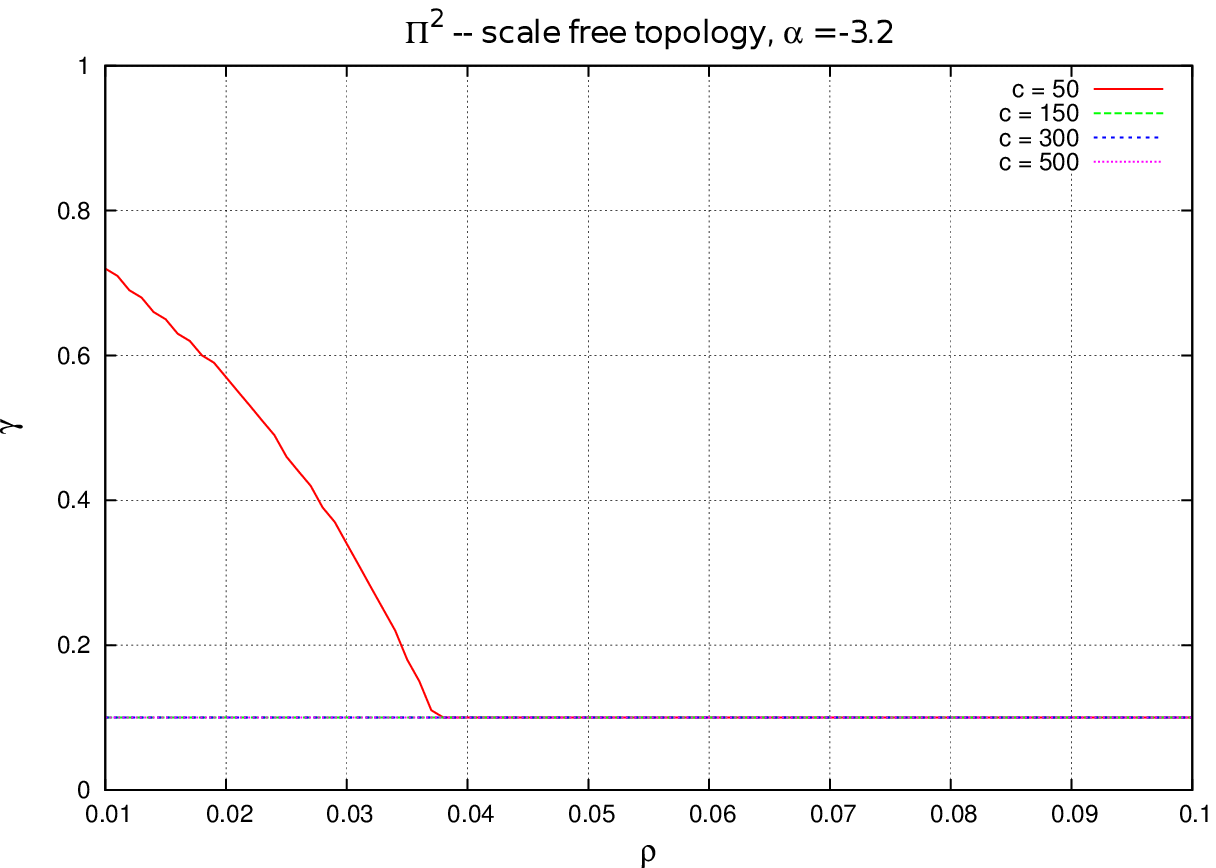}}
\caption{$\gamma$ value to obtain an infinite amount of query hits; scale-free network topologies with different power law distributions.}
   \label{fig:soglie}
\end{figure*}

It has been already mentioned that scale-free networks are characterized by the presence of hubs; moreover, we already mentioned the importance of introducing a cutoff that limits the maximum amount of contacts a peer may have in the overlay. Figure \ref{fig:soglie} shows the percolation transition values (i.e.~those values of $\gamma$ and $\rho$ above which queries do percolate through the net) for different scale-free networks, when varying the exponent $\alpha$ of the degree distribution\footnote{In this case, the cutoff imposes a limit on the moments of the degrees, that do not diverge; hence, it is interesting to consider networks with values of $\alpha$ higher than those considered above.} (different rows in the figure), the depth $k$ of the $k$-neighborhood (different charts in each row), and different settings for the cutoff $c$ (different curves on each chart).
Results are obtained through numerical measurements exploiting the analytical model. 
In this case, the cutoff has an influence on the ability of nodes to disseminate the query. In fact, the lower the cutoff the lower the number of links leaving from the hubs, and thus the more difficult is to spread the query. 
An interesting result related to the introduction of the cutoff, in line with what already mentioned, is that the lower the exponent $\alpha$ of the power law distribution, the higher the $\gamma$ to let queries percolate. 
This is due to the fact that the presence of the cutoff avoids that the first and second moments of the degree diverge. Moreover, the lower the exponent $\alpha$ the faster the distribution goes to $0$, and thus the higher the probability that nodes have low degrees, and thus the lower the connectivity of the network and its ability to spread contents.

Similarly, and as expected, in Figure \ref{fig:soglie} the higher the cutoff the lower the $\gamma$ to let queries percolate, since the presence of nodes with higher degrees (hubs) augments the connectivity of the network and its ability to spread contents.

Of course, when nodes have knowledge of $2$-neighbors, very small $\gamma$ values are needed with lower cutoffs (see charts on the right in the figure), while negligible values of $\gamma$ are necessary for higher settings of the cutoff $c$.

To sum up, outcomes confirm that lookup operations can be easily built over scale-free unstructured overlays.

\section{Conclusions}\label{sec:conc}

We analyzed the performance of a class of simple dissemination protocols, employing local knowledge of peers' neighborhood and gossip, to perform resource lookup over \ac{P2P} unstructured overlays. 
The provided analytical framework allows to tune the gossip probability to spread queries through the overlay, given a network topology and a resource probability distribution. These network parameters can be estimated using some techniques such as entropy-reduction protocols \cite{montresorMJB04}.

We tested our approach over scale-free networks.
It turns out that, in certain scenarios, it might be difficult to locate rare items with naive informed schemes without gossip (especially if $\Pi^1$ is exploited); this is in accordance with some previous results \cite{PuttaswamySZ08}. However, in most cases very low gossip probabilities are sufficient.
Thus, when networks are large in size and with a high level of churn, these solutions represent an interesting alternative to dissemination strategies built on top of costly structured distributed systems.

%
%


\begin{thebibliography}{10}

\bibitem{Aiello00arandom}
W.~Aiello, F.~Chung, and L.~Lu.
\newblock A random graph model for power law graphs.
\newblock {\em Experimental Math}, 10:53--66, 2000.

\bibitem{Cholvi:2004}
V.~Cholvi, P.~Felber, and E.~Biersack.
\newblock Efficient search in unstructured peer-to-peer networks.
\newblock In {\em Proc.~of the 16th ACM symposium on Parallelism in algorithms
  and architectures}, SPAA '04, pages 271--272, New York, NY, USA, 2004. ACM.

\bibitem{simutools}
G.~D'Angelo and S.~Ferretti.
\newblock Simulation of scale-free networks.
\newblock In {\em Simutools '09: Proc.~of the 2nd International Conference on
  Simulation Tools and Techniques}, pages 1--10, ICST, Brussels, Belgium, 2009.
  ICST.

\bibitem{disio11}
G.~D'Angelo, S.~Ferretti, and M.~Marzolla.
\newblock Adaptive event dissemination for peer-to-peer multiplayer online
  games.
\newblock In {\em Proc.~of the Int.~Conf.~on Simulation Tools and Techniques
  (SIMUTools 2011)}. ICST, 2011.

\bibitem{ferretti_trans.cs.2012.10-12.e2}
S.~Ferretti.
\newblock On the degree distribution of faulty peer-to-peer overlay networks.
\newblock {\em EAI Endorsed Transactions on Complex Systems}, 12(1), 11 2012.

\bibitem{simplex}
S.~Ferretti.
\newblock Publish-subscribe systems via gossip: a study based on complex
  networks.
\newblock In {\em Proc.~of the 4th Annual Workshop on Simplifying Complex
  Networks for Practitioners}, SIMPLEX '12, pages 7--12, New York, NY, USA,
  2012. ACM.

\bibitem{Hidalgo:2011}
N.~Hidalgo, E.~Rosas, L.~Arantes, O.~Marin, P.~Sens, and X.~Bonnaire.
\newblock Dring: A layered scheme for range queries over dhts.
\newblock In {\em Proc.~of the 2011 IEEE 11th International Conference on
  Computer and Information Technology}, CIT '11, pages 29--34. IEEE, 2011.

\bibitem{Keidar:2006}
I.~Keidar and R.~Melamed.
\newblock Evaluating unstructured peer-to-peer lookup overlays.
\newblock In {\em Proceedings of the 2006 ACM symposium on Applied computing},
  SAC '06, pages 675--679, New York, NY, USA, 2006. ACM.

\bibitem{montresorMJB04}
A.~Montresor, M.~Jelasity, and O.~Babaoglu.
\newblock Robust aggregation protocols for large-scale overlay networks.
\newblock In {\em Proc. of the 2004 Int. Conference on Dependable Systems and
  Networks (DSN'04)}, pages 19--28, Florence, Italy, June 2004. IEEE Computer
  Society.

\bibitem{newmanHandbook}
M.~E.~J. Newman.
\newblock {\em Random graphs as models of networks}, pages 35--68.
\newblock Wiley-VCH Verlag GmbH and Co. KGaA, 2005.

\bibitem{PuttaswamySZ08}
K.~P.~N. Puttaswamy, A.~Sala, and B.~Y. Zhao.
\newblock Searching for rare objects using index replication.
\newblock In {\em 27th IEEE International Conference on Computer
  Communications}, pages 1723--1731. IEEE, 2008.

\bibitem{voulgaris.jnsm.2005}
S.~Voulgaris, D.~Gavidia, and M.~{van Steen}.
\newblock Cyclon: Inexpensive membership management for unstructured p2p
  overlays.
\newblock {\em Journal of Network and Systems Management}, 13(2):197--217, June
  2005.

\end{thebibliography}
\end{document}